\documentclass[
]{ceurart}

\usepackage{booktabs}

\begin{document}

\copyrightyear{2021}
\copyrightclause{Copyright for this paper by its authors.
  Use permitted under Creative Commons License Attribution 4.0
  International (CC BY 4.0).}

\conference{Forum for Information Retrieval Evaluation, December 13-17, 2021, India}

\title{A Semantic Search Pipeline for Causality-driven Adhoc Information Retrieval}

\author[1]{Dhairya Dalal}[%
orcid=0000-0003-0279-234X,
email=d.dalal1@nuigalway.ie]
\address[1]{SFI Centre for Research and Training in Artificial Intelligence, Data Science Institute, National University of Ireland Galway}

\author[2]{Sharmi Dev Gupta}[%
orcid=0000-0003-0376-2206,
email=sharmi.devgupta@cs.ucc.ie,
]
\address[2]{SFI Centre for Research and Training in Artificial Intelligence, 
	     School of Computer Science and Information Technology, University College Cork}

\author[1]{Bentolhoda Binaei}[%
orcid=,
email=b.binaei1@nuigalway.ie,
]

\begin{abstract}
We present a unsupervised semantic search pipeline for the Causality-driven Adhoc Information Retrieval (CAIR-2021) shared task. The CAIR shared task expands traditional information retrieval to support the retrieval of documents containing the likely causes of a query event. A successful system must be able to distinguish between topical documents and documents containing causal descriptions of events that are causally related to the query event. Our approach involves aggregating results from multiple query strategies over a semantic and lexical index. The proposed approach leads the CAIR-2021 leaderboard and outperformed both traditional IR and pure semantic embedding-based approaches.
\end{abstract}



\maketitle

\section{Introduction}
\begin{figure}[htp]
    \centering
    \includegraphics[width=10cm]{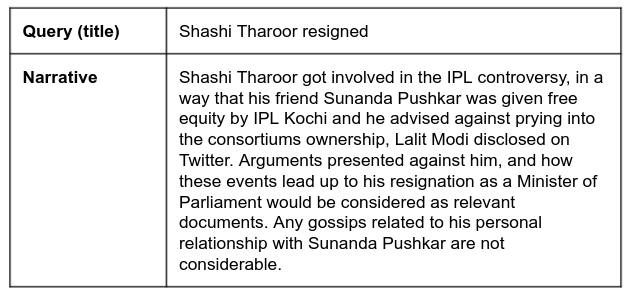}
    \caption{Example CAIR topic. Each topic consists of a query (title), which describes an event, and a narrative, which contains descriptions of documents that are causally relevant to the event.}
    \label{fig:example}
\end{figure}

The Causality-driven Adhoc Information Retrieval (CAIR) shared task consists of retrieving documents with the likely causes of a query event \cite{cairoverview}. The search system must be able to differentiate between topical documents and casual documents. Traditional information retrieval (IR) systems usually rely on keyword matching and corpus level n-gram statistics to score which documents are most topically relevant to a provided query. In contrast, given a query event (e.g. Shashi Tharoor resigned), the goal of the causal search system is to identify documents that contain causal information about the events that lead to the query event. For example, causally relevant documents for the query in Figure \ref{fig:example} would refer to the IPL controversy and illicit behavior by Shashi Tharoor. General documents that mention Shashi Tharoor, while topically relevant, may not be causally relevant if they do not contain information about his misbehavior. 

In this paper, we describe our solution for the CAIR shared task. We design a unsupervised semantic search pipeline, which aggregates results across several query strategies and indices. The pipeline leverages both a lexical index and a semantic index to retrieve causally relevant documents. Our approach both outperformed standard IR baselines and semantic baselines and was the top method on the CAIR-2021 task leaderboard. 

\section{Related Works}
\citet{Datta2020WheresTW} provide a brief survey of the literature on causality in natural language processing and explore the task of causal information retrieval in the context of news articles. They also introduce a recursive causal retrieval model which allows for identifying the causal chain of events that led to a news event. \citet{datta2020retrieving} propose an unsupervised pseudo-relevance feedback approach that estimates the distribution of infrequent terms that are potentially relevant to the causality of the query event. Recent advances in IR have focused on neural re-ranking and leveraging latent embeddings to improve the overall recall and semantic relevance of returned results. For example, \citet{pang2020setrank} propose SetRank, a permutation-invariant ranking model that jointly learns the embeddings of retrieved documents using self-attention. Most modern IR approaches combine lexical and semantic approaches. For example, \citet{gao2021complement} presents CLEAR in which a residual-based learning framework teaches the neural embedding to be complementary to the lexical retrieval model.  Our approach follows the trend of combining lexical models with semantic embeddings. 

\section{Methods}
Our approach focused on developing an unsupervised semantic search pipeline. Documents were indexed in two indices: a semantic index and a lexical index (see Section \ref{section:index}). Results from multiple queries across the two indices were then aggregated to return the most relevant documents. We additionally explored a post query filter step that aimed to identify documents that contained causal language in the context of the query event. This approach did not produce viable results and was not pursued. In this section, we will present our methodology and experimental setup in further detail.  

\subsection{Document Indexing} \label{section:index}
Two document indices were created for our semantic search pipeline. The first was a \textbf{lexical index} that treated documents as bags of words and was optimized for Okapi BM25 \cite{Robertson1994OkapiAT} retrieval. Before indexing, documents were cleaned and tokenized using standard preprocessing steps: lowercasing, stripping out all non-alphanumeric characters, and lemmatization. Thus each document $D$ was broken into lemmatized unigram tokens $t_1 ... t_n$. Next, the tokenized documents were further processed to support the Okapi BM25 ranking algorithm. Given a query $Q$ which consists of query tokens $q_1, ... q_n$, we score each document D in our index using the following scoring function:

\begin{math}
Score(D,Q) = \sum_{i=1}^{n} IDF(q_i) \cdot \frac{freq(q_i, D) \cdot (k_1+1)}{freq(q_i, D) + k_1 \cdot (1 - b + b \cdot \frac{D_length}{avgDL})}
\end{math}

$IDF(q_i)$ is the inverse document frequency of the query token and $freq(q_i, D)$ is the frequency of query token in the document. Finally, $D_{length}$ is the length of the document (i.e. the total number of tokens) and $avgDL$ is the average document length. Okapi BM25 scores are unbounded and larger scores indicate the retrieved document is more relevant compared to lower scored documents in the context of the query. 

The second index was a \textbf{semantic index} where documents were represented fixed dimension vector embeddings generated by a sentence embedding model. The broad goal of the semantic index was to retrieve documents that are semantically similar to the query. Semantic relevance is measured by the cosine similarity between query embedding $U$ and document embedding $V$ which can be defined as: 

\begin{math}
similarity(U, V) = \frac{U\cdot V}{\left \| U \right \|\cdot  \left \|  V\right \|} = \frac{\sum_{i=0}^{n} U_{i} V_{i}}{\sqrt{\sum_{i=1}^{n}U^2_{i}} \sqrt{\sum_{i=1}^{n}V^2_{i}} }
\end{math}

\begin{figure}[htp]
    \centering
    \includegraphics[width=15cm]{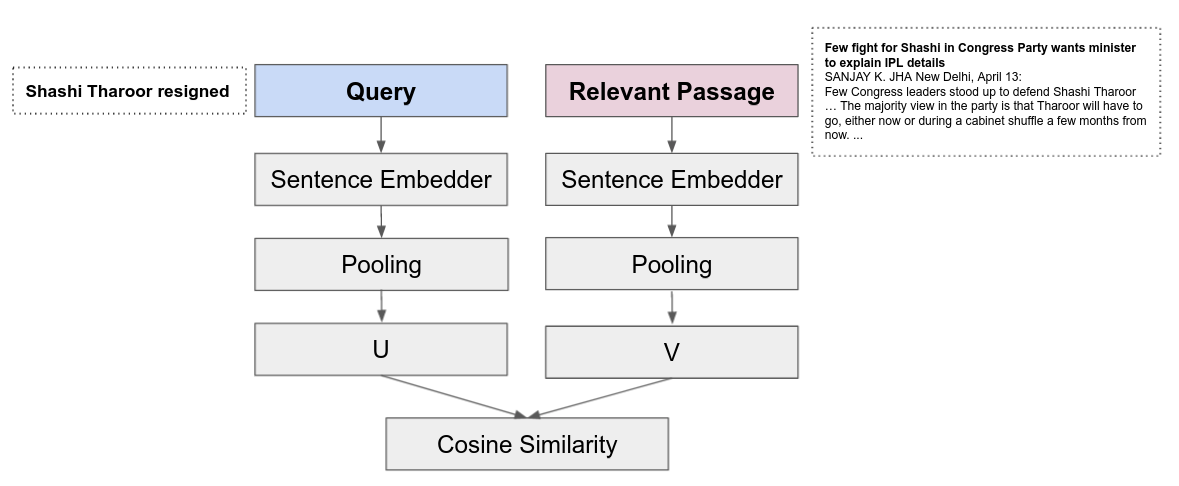}
    \caption{Siamese sentence embedding architecture for asymmetric matching.}
    \label{fig:siamese}
\end{figure}

Sentence embedding models generate a fixed dimension representation for a provided input text and are trained to represent sentence-level inputs for tasks such as semantic text similarity, paraphrase detection, and textual entailment. Search applications present challenges when considering sentence embedding models. Applying sentence embedding to document-level inputs (e.g. paragraphs or new articles) dilutes the quality of embedding representation and will likely result in poorer performance in the context of dense passage retrieval and ranking. Additionally, there is an input asymmetry challenge where the query length is often shorter than the relevant document that is to be retrieved. Finally, there may be limited lexical overlap between the query text and the relevant document. As a result standard sentence embeddings models like USE (Universal Sentence Encoder) \cite{cer-etal-2018-universal} will struggle for general semantic search use cases. To account for this we use a Siamese network architecture \cite{reimers-2019-sentence-bert} that was pretrained to support asymmetric  (Figure \ref{fig:siamese}) matching. The Siamese architecture takes as input query and relevant passage pairs and fine-tunes a shared sentence embedding model to increase the cosine similarity between relevant pairs and decreases the similarity between negative pair samples. The resulting sentence embedding model is better tuned to support the asymmetric nature of determining the semantic similarity between a query and document embedding. Details on the pretrained sentence embedding model can be found in Section \ref{sec:setup}.

\subsection{Semantic Search Pipeline}

\begin{figure}[htp]
    \includegraphics[width=15cm]{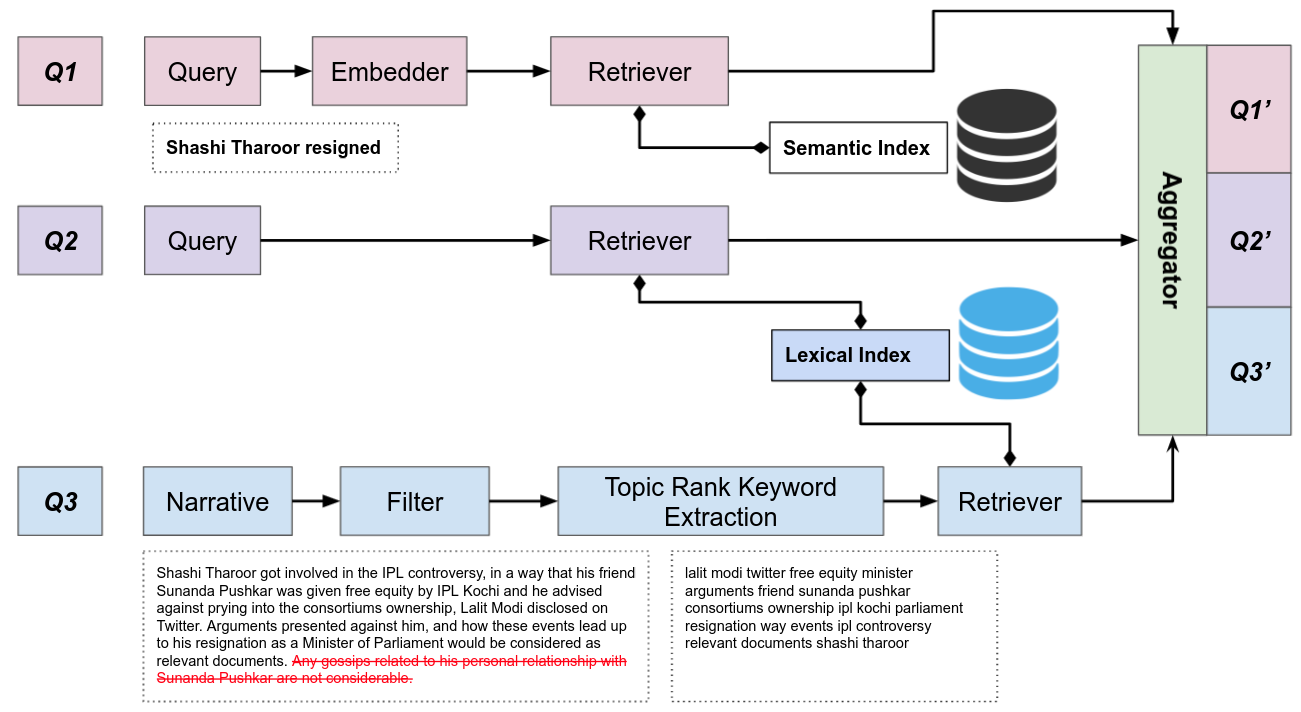}
    \caption{The semantic search pipeline aggregates results from three query strategies, $Q1$, $Q2$, and $Q3$. $Q1$ embeds the query using the sentence embedding model and retrieves the most relevant results based on cosine similarity. $Q2$ and $Q3$ retrieve the most relevant documents from the lexical index. $Q3$ adds filtering and keyword extraction steps to transform the narrative description in causal search terms. Finally results from all three queries ($Q1'$, $Q2'$, and $Q3'$) are aggregated and re-ranked by the aggregator module. The top 500 relevant submissions are returned.}
    \label{fig:pipeline}
\end{figure}

Our semantic search pipeline (Figure \ref{fig:pipeline}) aggregates results from three distinct query strategies to produce the final set of relevant causal documents. Provided a topic consisting of a title and narrative (e,g, Figure \ref{fig:example}), we treat the title as the query text and narrative as a source for causal keywords. 

$Q1$ retrieves the 500 most semantically similar documents from the semantic index. This is accomplished by embedding the query text using the sentence embedding model, retrieving the closest document embeddings based on cosine distance, and then ranking the documents using cosine similarity scores between query embedding and document embedding. 

$Q2$ retrieves the 500 most relevant documents from the lexical index, where the relevance is measured by the Okapi BM25 between the candidate documents tokens and query tokens.

$Q3$ also retrieves 500 results from the lexical index but uses causal keywords extracted from the narrative description. The narrative text is first passed through a filter step which removes any statements in the description that describes irrelevant documents. The filter uses a simple keyword-based regex (e.g. not relevant, not considered, irrelevant, etc) to identify those statements. Next, the filtered narrative is converted into a set of keywords using TopicRank \cite{bougouin-etal-2013-topicrank}. Finally, the causal keywords from the narrative are used to query the lexical index. 

$Q1$, $Q2$, and $Q3$ each produce a set of candidate documents ($Q1'$, $Q2'$, and $Q3' $ respectively). These results are sent to an aggregator module that deduplicates and re-ranks all the candidate documents. If a document appears in multiple results sets, its scores are summed. The top 500 documents are returned as the final result set.

\subsection{Post Query Causal Filtering}
We additionally explored a post query filtering step. This involved extracting causal relations (if any were found) from the candidate document. Candidate documents would have passed this filtering stage if the extracted cause had an overlap with the query text and the extracted effect overlapped with the narrative causal keywords. This approach did not yield promising results on the train topics and was not explored further on the test topics. Often the causal documents did not mention the caused event as the document was reporting news that occurred before the query event. This filtering method would have failed to identify news reports of events in the past that lead to the query event because at the time of the reporting, the article did not know about the query event (as it would happen in the future). 

\section{Experiments}
In this section, we describe our implementation and experiment results. 

\subsection{Data}
The CAIR dataset contains 303,291 Telegraph India news articles from 2001 to 2010 \cite{datta2020retrieving}. There are 5 train topics and 20 test topics provided. Each topic (e.g. Figure \ref{fig:example}) consists of a title, which describes the query event, and a narrative that describes the expected relevant and irrelevant documents.

\subsection{Setup}
\label{sec:setup}
The spacy library \footnote{https://spacy.io} was used for prepossessing (i.e. lemmatizing and tokenizing). We used the python rank25 library \footnote{https://github.com/dorianbrown/rank\_bm25} to implement a lexical index optimized for Okapi BM25 scoring. The default values were used for the $k1$ (1.5) and $b$ (0.75) parameters. 

For the semantic index, we use the pretrained \textit{msmarco-distilbert-base-v4} sentence embedding model from the SentenceTransformers library \cite{reimers-2019-sentence-bert}. This model was pretrained on the MS Marco passage ranking dataset \cite{msmarco} which has asymmetric input properties as the query is often shorted the relevant passage. The MS Marco dataset consists of a million queries from the Bing search engine and 8.8 million passages from search results. The passage ranking task requires the model to find the most relevant passages for a provided query and rank them. Documents and qrels from the CAIR corpus were not used for the pretraining of the sentence embedding model. 

All the documents in the CAIR corpus were embedded using the  \textit{msmarco-distilbert-base-v4} sentence embedding model and then stored in an index optimized for approximate nearest neighbors search. We used the ANNOY python library \footnote{https://github.com/spotify/annoy} to store the document embeddings and built a search index of 1000 trees.

\subsection{Baselines}
We evaluated our approach against four different lexical and semantic baselines. All the baselines returned the top 500 relevant results which were evaluated against the gold document relevance set. Mean Average Precision (MAP) and Precision at 5 (P@5) metrics were used for evaluation. The first (Narrative Only Okapi BM25) baseline used returned results from the lexical index using the narrative text as the query. The second baseline (Query Only Okapi BM25) used the title as the query for lexical index. The third baseline (Query + Narrative Semantic) combined the query and title texts and retrieved the most relevant semantic results from the semantic index. Finally, the last baseline only used the title text to query the semantic index. 

\subsection{Results}

Experiment results can be found in Table \ref{table:results}. In addition to our baselines, we include the results of the best submission from the NITS team in the CAIR 2021 shared task. The test set contained 20 topics and a gold relevance set which identified causally relevant documents in the corpus. Our semantic search pipeline outperforms all the baseline methods and leads the shared task leader board. The semantic search pipeline posted a twenty-five percent increase in MAP and a fourteen percent increase in P@5 over the Narrative Only Okapi BM25 baseline. 

\label{table:results}
\begin{table}[!htp]
\caption{Experiment Results for the 20 test topics.}
\begin{tabular}{@{}lll@{}}
\toprule
\textbf{Method}                    & \textbf{MAP} & \textbf{P@5} \\ \midrule
Semantic Pipeline                  & .5761        & .7800        \\
Narrative Only Okapi BM25 baseline  & .3285        & .6399        \\
Query Only Okapi BM25 baseline     & .2561        & .4999        \\
Query + Narrative Semantic baseline & .2239        & .5500        \\ 
Query Only Semantic baseline       & .1611        & .5000        \\ 
NITS-Run                           &.1063	      &.4800          \\\bottomrule
\end{tabular}

\end{table}

Our semantic search pipeline uses the same lexical and semantic indexes as the baselines. However, the pipeline is better able to combine the lexical and semantic results to produce the most causally relevant documents. The aggregator module conceptually functions as an ensemble model and weights documents that appear in multiple query result sets higher. Each query strategy utilized information from the topic differently and the final result set reflected that. 

\begin{figure}[htp]
    \centering
    \includegraphics[width=14cm]{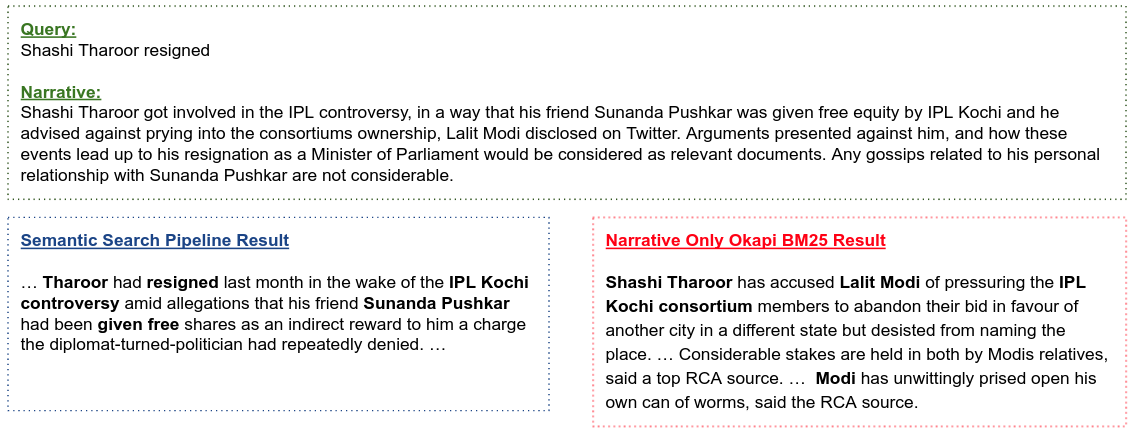}
    \caption{Example results returned by Semantic Search Pipeline and the Narrative Only Okapi BM25 baseline. The baseline returns a topically relevant result based on keyword matches but fails to describe why Shashi Tharoor resigned.}
    \label{fig:error}
\end{figure}

Amongst the baselines, the Narrative Only Okapi BM25 baseline was the strongest. The narrative text contains the most useful information about what caused the query event and was expected to provide the best results amongst the baselines. However, the narrative input with a lexical index is still prone to returning topical documents that are not causally relevant. Figure \ref{fig:error} provides a qualitative comparison between the Narrative Only Okapi BM25 baseline and the Semantic Search Pipeline. The baseline models match on terms present in the narrative but the article is focused on accusing Modi of misconduct in the context of the IPL Kochi scandal. In contrast, the Semantic Search Pipeline correctly identifies a document that describes why Shashi Tharoor resigned in relation to scandal and his friend Sunanda Pushkar.

\section{Conclusion}
In this paper, we introduced a semantic search pipeline for the CAIR-2021 shared task. Our approach aggregated results from multiple query strategies across a lexical and semantic index. The semantic search pipeline outperformed the lexical and simple semantic baselines and was the top method on the CAIR 2021 leader board. This approach should serve as a stepping stone toward better causal information retrieval. Future work could explore developing a better model of causality and retrieving results using the query title only. The narrative text provides strong clues as the causal terms that would be in the causally relevant documents. A causal search system would have a better way identify and causally linking events. 

\begin{acknowledgments}
This work was supported by Science Foundation Ireland under grants SFI/18/CRT/6223 (Centre for Research Training in Artificial Intelligence).
\end{acknowledgments}

%
\bibliography{sample-ceur}

\appendix

\end{document}